# SYMMETRIC AND ANTI-SYMMETRIC QUANTUM FUNCTIONS

J. Robert Burger
California State University Northridge
CSUN Report Number HCEEN006_2A

**ABSTRACT**
This paper introduces and analyzes symmetric and anti-symmetric quantum binary functions. Generally, such functions uniquely convert a given computational basis state into a different basis state, but with either a plus or a minus sign. Such functions may serve along with a constant function (in a Deutsch-Jozsa type of algorithm) to provide 2**n deterministic qubit combinations (for n qubits) instead of just one.

**BACKGROUND**
The quantum functions that are the subject of this paper are noteworthy because they are especially easy to work with, and because they relate to the ideas of Deutsch, Jozsa and others [1, 2, 3]. Symmetric and anti-symmetric quantum binary functions are first cousins to symmetric and anti-symmetric Boolean functions. Starting with $|x_1>$, $|x_2>$, $|x_3>$…, $|x_{n+1}>$ where each qubit $|x_i>$ is either $|0>$ or $|1>$, one may construct $|Hx_1>$, $|Hx_2>$, $|Hx_3>$…, $|Hx_{n+1}>$ where **H** is the Hadamard matrix, defined to be $\begin{bmatrix} 1 & 1 \\ 1 & -1 \end{bmatrix}$. These qubits can form a state vector, denoted when the commas are replaced with "direct" product symbols $\otimes$, for example: $|Hx_1> \otimes |Hx_2> \otimes |Hx_3> \ldots \otimes |Hx_{n+1}>$ or sometimes simply: $|Hx_1, Hx_2, Hx_3,\ldots, Hx_{n+1}>$.

A quantum function is fundamentally a unitary matrix that operates on a state vector. A symmetric or anti-symmetric quantum binary function will achieve $\pm |Hy_1> \otimes |Hy_2> \otimes |Hy_3> \ldots \otimes |Hy_{n+1}>$ where each qubit $|y_i>$ is either $|0>$ or $\pm|1>$; this form is easily factored to give $\pm|y_2>, |y_3>\ldots, |y_{n+1}>$, where each qubit is known with certainty to be either $|0>$ or $\pm|1>$.

The main purpose of this work is to explore the interesting properties of a certain fundamental and well-defined class of quantum binary functions. Obviously one could apply NOT gates to selected input qubits to accomplish most of what a symmetric or anti-symmetric quantum binary function does to the output qubits. But there are minor complexities, such as outputting a negative state vector when inputting a positive state vector.

**SYMMETRIC AND ANTI-SYMMETRIC FUNCTIONS**
Often in quantum algorithms a "superposition" is constructed beginning with all $|x_i> = |0>$ for i from 1 to n, and often with $|x_{n+1}> = |1>$. In contrast, this paper considers a random assortment of initial $|x_i>$ each of which may be either $|0>$ and $|1>$. Consider now a quantum machine with $|x_{n+1}> = |1>$ as in Figure 1.

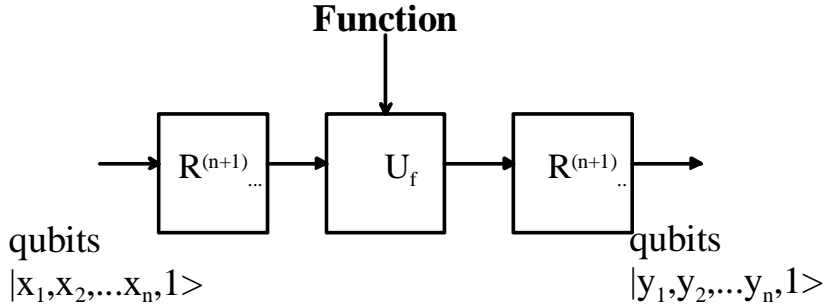

**Figure 1. Basic Quantum Machine**

qubits $|x_1, x_2, ... x_n, 1\rangle$ ... qubits $|y_1, y_2, ... y_n, 1\rangle$

$R^{(n+1)}$ refers to n+1 applications of **H**, that is: $H \otimes H \otimes ... \otimes H$, where $\otimes$ denotes a "direct" matrix product. The qubit $|x_{n+1}\rangle$ initialized to $|1\rangle$ and is needed to store the value of the function. $U_f$ denotes a unitary square matrix of dimension n+1 and is assumed to represent a binary function of interest; a suitable $U_f$ with real values may be denoted by: $U_f |i, j, k\rangle = |i, j, k \oplus f(i, j)\rangle$ where $\oplus$ is the logical exclusive OR. The $|i\rangle$, $|j\rangle$, and $|k\rangle$ refer to qubit "computational" basis states, that is, each is $|0\rangle$ or $|1\rangle$. To be a "computational" basis state, each qubit is either a zero $|0\rangle = [1\ 0]'$ or a one $|1\rangle = [0\ 1]'$, sometimes pictured as spin up or spin down. The prime denotes transpose. So the basic states are: $|0\rangle = \begin{bmatrix} 1 \\ 0 \end{bmatrix}$; $|1\rangle = \begin{bmatrix} 0 \\ 1 \end{bmatrix}$. With considerable imagination, it is possible to construct a diagonal-like $U_f$ from this mysterious definition. For example, $U_f$ for the AND function is:

$$U_f = \begin{bmatrix} 1 & 0 & 0 & 0 & 0 & 0 & 0 & 0 \\ 0 & 1 & 0 & 0 & 0 & 0 & 0 & 0 \\ 0 & 0 & 1 & 0 & 0 & 0 & 0 & 0 \\ 0 & 0 & 0 & 1 & 0 & 0 & 0 & 0 \\ 0 & 0 & 0 & 0 & 1 & 0 & 0 & 0 \\ 0 & 0 & 0 & 0 & 0 & 1 & 0 & 0 \\ 0 & 0 & 0 & 0 & 0 & 0 & 0 & 1 \\ 0 & 0 & 0 & 0 & 0 & 0 & 1 & 0 \end{bmatrix}$$

Generalization to larger n for any Boolean function is straightforward once this definition of $U_f$ is deciphered. The input $|x_1, x_2, 1\rangle$ denotes $|x_1\rangle \otimes |x_2\rangle \otimes |1\rangle$, that is, a vector with eight elements numbered 0 to 7. For example, $|0, 0, 1\rangle$ means $[0\ 1\ 0\ 0\ 0\ 0\ 0\ 0]'$.

Table 1 shows truth tables for symmetric and anti-symmetric binary functions of two bits a, b. Each function in the table is defined as a, b steps up from zero, for instance: 00, 01, 10, 11. A given truth table may be delineated in a row instead of a column. For example, 0001 would represent the AND function. But the AND does not appear in the table because it is neither symmetric nor anti-symmetric.

Definition: A symmetric binary function has a truth table that is mirrored about its center. For instance, the function $f_6$ has a truth table that can be delineated as 0110. The center is between the two ones. The first two bits 0, 1 are mirrored by the last two bits 1, 0. Thus 0110 is defined here to be symmetric.

Definition: An anti-symmetric binary function has a truth table in which the bit-wise complement of the first half is mirrored in the second half. Thus the function $f_5$ which is 0101 is anti-symmetric. The first two bits 01 are mirrored by the complements of the last two bits 0'1' or 10.

In the quantum machine of Figure 1, a symmetric or anti-symmetric function results in a computational basis state at the output. The $\psi_i$ in Table 1 are the elements of the output state vector of dimension n+1 for n = 2.

**Table 1. Symmetric Anti-symmetric Truth Tables & State Vectors n = 2**

| ab | $f_0$ | $f_3$ | $f_5$ | $f_6$ | $f_9$ | $f_A$ | $f_C$ | $f_F$ |
|---|---|---|---|---|---|---|---|---|
| 00 | 0 | 0 | 0 | 0 | 1 | 1 | 1 | 1 |
| 01 | 0 | 0 | 1 | 1 | 0 | 0 | 1 | 1 |
| 10 | 0 | 1 | 0 | 1 | 0 | 1 | 0 | 1 |
| 11 | 0 | 1 | 1 | 0 | 1 | 0 | 0 | 1 |
| $\psi_0=$ | 0 | 0 | 0 | 0 | 0 | 0 | 0 | 0 |
| $\psi_1=$ | 1 | 0 | 0 | 0 | 0 | 0 | 0 | -1 |
| $\psi_2=$ | 0 | 0 | 0 | 0 | 0 | 0 | 0 | 0 |
| $\psi_3=$ | 0 | 0 | 1 | 0 | 0 | -1 | 0 | 0 |
| $\psi_4=$ | 0 | 0 | 0 | 0 | 0 | 0 | 0 | 0 |
| $\psi_5=$ | 0 | 1 | 0 | 0 | 0 | 0 | -1 | 0 |
| $\psi_6=$ | 0 | 0 | 0 | 0 | 0 | 0 | 0 | 0 |
| $\psi_7=$ | 0 | 0 | 0 | 1 | -1 | 0 | 0 | 0 |

There is a simple relationship between the truth table entries and the unitary matrix. A zero in the truth table results in a sub-matrix $\begin{bmatrix} 1 & 0 \\ 0 & 1 \end{bmatrix}$ on the diagonal of the matrix; a one in the truth table results in a sub-matrix $\begin{bmatrix} 0 & 1 \\ 1 & 0 \end{bmatrix}$ on the diagonal. Thus $\mathbf{U_f}$ has a sub-structure that matches the truth table.

Additional examples follow: Starting with $|0, 0, 1\rangle$, the first $\mathbf{R^{(3)}}$ in Figure 1 results in (1 -1 1 -1 1 -1 1 -1) /√8. The first four functions of, $f_1$, $f_3$, $f_5$, $f_6$ in Table 1 transform this to (1 -1 1 -1 1 -1 1 -1) /√8, (1 -1 1 -1 -1 1 -1 1) /√8, (1 -1 -1 1 1 -1 -1 1) /√8, or (1 -1 -1 1 -1 1 1 -1) /√8. Note that elements are being interchanged, so states could conveniently be represented with cyclic notation, as done in reversible design [5, 6], although such design is unnecessary here. We simply note that the state vector has pairs interchanged whenever there is a '1' in the truth table.

Symmetry and anti-symmetry are must useful when factoring state vectors into qubit components. It allows $|\mathbf{Hy_1,Hy_2,Hy_3…Hy_n}\rangle$ to be factored into $|\mathbf{Hy_1}\rangle |\mathbf{Hy_2}\rangle |\mathbf{Hy_3}\rangle$

...|**Hy$_n$**>.  The vectors in the previous paragraph each can be factored into the form ± (a b)/√2 ⊗ (c d)/√2 ⊗ (e f)/√2 where each of a, b, c, d, e, f are either 1 or –1.  Consider for example (1 -1 -1 1 -1 1 1 -1) /√8.  It has overall anti-symmetry so |**Hy$_3$**> = (1 –1)/√2. Each half (or foursome) also has anti-symmetry; so **Hy$_2$**> = (1 –1)/√2.  Finally, each half of each half (that is, each twosome) has anti-symmetry; so **Hy$_1$**> = (1 –1)/√2.  Applying **H** to each factored qubit, it can be concluded that |**y$_1$**> = |**1**>, |**y$_2$**> = |**1**>, and |**y$_3$**> =|**1**>. The sign of the qubits cannot be determined by qubit readout since all that may be observed are magnitudes.

Returning to Table 1, the complements of the functions $f_9$, $f_A$, $f_C$, $f_F$ provide a negative sign in the resulting vector.
Definition:  Negative symmetric or anti-symmetric quantum functions are those which give a negative sign in the computational basis vector.

For example, the function 0000 gives [0 1 0 0 0 0 0 0]' while the complement function 1111 gives [0 -1 0 0 0 0 0 0]'.  Negative states can thus be created by symmetric or anti-symmetric functions and might be required internally, although they cannot be observed directly.

**CONSTRUCTION OF SYMMETRIC AND ANTI-SYMMETRIC FUNCTIONS**
Functions of interest can be constructed as in Appendix 1.  It turns out that there are a total of $2^{n+1}$ symmetric or anti-symmetric (positive and negative) binary functions on n bits.  This may be compared to the common knowledge that there are a grand total of $2^{2^n}$ binary functions for n bits.

**BOOLEAN ALGEBRAIC FORMS**
Examples of Boolean versions of positive functions are 0, $p_1$, $p_2$, and $p_1 \oplus p_2$ for n = 2 as in Table 2.  Negative functions are the Boolean complements.

**Table 2. Positive functions for n=2**

| f($p_1$,$p_2$) | bin | hex | dec |
|---|---|---|---|
| 0 | 0000 | 0 | 0 |
| $p_1$ | 0011 | 3 | 3 |
| $p_2$ | 0101 | 5 | 5 |
| $p_1 \oplus p_2$ | 0110 | 6 | 6 |

Table 3 list the functions for n = 3.  Note that the components in Table 3 are 0, $p_1$, $p_2$, and $p_3$, the XOR of combinations of $p_1$,$p_2$,$p_3$ taken 2 at a time, and the XOR of all 3.  The total number of positive functions is eight, or twice as many as in Table 2.  Following this reasoning as n increases, it can be shown that there are $2^n$ positive and $2^n$ negative, or $2^{n+1}$ positive and negative functions, which agrees with the number in the appendix.

**Table 3. Positive functions for n=3**

| f($p_1$,$p_2$,$p_3$) | bin | hex | dec |
|---|---|---|---|
| 0 | 00000000 | 00 | 0 |
| $p_1$ | 00001111 | 0F | 15 |
| $p_2$ | 00110011 | 33 | 21 |
| $p_1 \oplus p_2$ | 00111100 | 3C | 60 |
| $p_3$ | 01010101 | 55 | 85 |
| $p_1 \oplus p_3$ | 01011010 | 5A | 90 |
| $p_2 \oplus p_3$ | 01100110 | 66 | 102 |
| $p_1 \oplus p_2 \oplus p_3$ | 01101001 | 69 | 105 |

**SHORTCUT METHOD FOR SYMMETRIC AND ANTI-SYMMETRIC FUNCTIONS**

Wiring diagrams have a special utility for symmetric and anti-symmetric quantum functions. Figure 3 part (a) is a wiring diagram for the function $f(p_1, p_2) = p_1$. Part b is an equivalent as suggested by Pittenger on page 28 [1]. Part b quickly determines the effect of part a. For examples |**0,0,1**> and $f(p_1, p_2) = p_1$ in Figure 3 becomes |**1,0,1**>; |**0,0,1**> and $f(p_1, p_2) = p_1 \oplus p_2$ in Figure 4 becomes |**1,1,1**>. Positive symmetric and anti-symmetric quantum functions effectively apply NOT gates to qubits, so wiring diagrams are useful tools here.

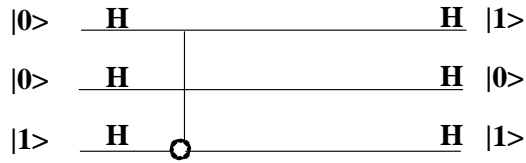

(a) f(p1,p2)=p1

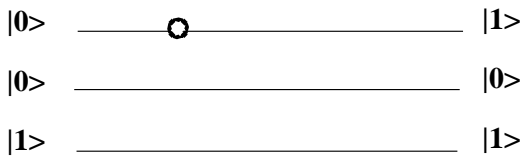

(b) Equivalent

**Figure 3. Equivalents for f(p1, p2) = p1**

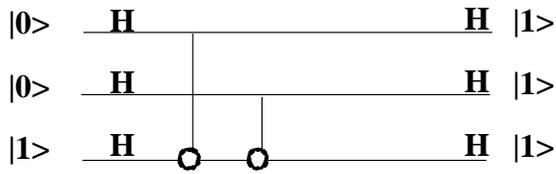

**(a) f(p1,p2)=XOR(p1,p2)**

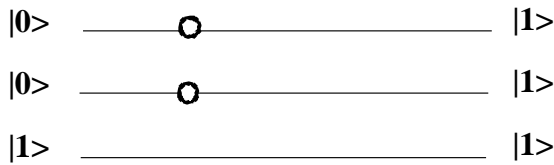

**(b) Equivalent**

**Figure 4 Equivalents for f(p1, p2) = p1⊕ p2**

Using these tools, a quantum machine in the form of Figure 1 may be <u>designed</u> to convert any input computational basis state into a different output computational basis state. Complementary and other functions are more involved. If a negative sign is required in a computational basis state vector, one may apply a negative symmetric or anti-symmetric quantum function.

**IDENTIFICATION OF FUNCTIONS**
When the $|x_i\rangle$ and the $|y_i\rangle$ are known, one of $2^n$ positive functions can be identified using the wiring diagram as a guide. First in part (b) the NOT gates are determined; then for part (a) the function is determined. To demonstrate this for larger n, consider positive functions for n = 4 as listed in Table 4 (constructed in analogy to Table 3).

Table 5 (in Appendix 2) is part of a larger table that was prepared to show how input computational basis states transform into given output computational basis states. It illustrates that for a given input, there is a one-to-one relationship between the function used and the output given.

Table 5 has a definite structure.
  A) The upper half of the positive functions a through h listed in Table 4 are located on negative slope diagonals. The lower half functions i through p are located on positive slope diagonals and are mirror images of the upper half.
  B) Functions a, b, c, …, p alternate between upper and lower halves of the table (refer to column 1).

Note that such tables become very large for large n.

**Table 4  Positive Functions n = 4**

| id | Hex ($) | dec |
|---|---|---|
| a | 0000 | 0 |
| b | 00FF | 255 |
| c | 0F0F | 3855 |
| d | 0FF0 | 4080 |
| **e** | **3333** | 13107 |
| f | 33CC | 13260 |
| **g** | **3C3C** | 15420 |
| h | 3CC3 | 15555 |
| i | 5555 | 21845 |
| j | 55AA | 21930 |
| k | 5A5A | 23130 |
| l | 5AA5 | 23205 |
| m | 6666 | 26214 |
| n | 6699 | 26265 |
| o | 6969 | 26985 |
| p | 6996 | 27030 |

**Design Problem A** – The input $|x,1\rangle$ is given as $|1,0,0,0,1\rangle$ and the desired output is $|1,1,1,0,1\rangle$.  Identify the positive quantum function.

**Solution A** – Note that $|x_2\rangle$ and $|x_3\rangle$ must be inverted, as may be seen from a wiring diagram applied as above.  Therefore, $f(p_1, p_2, p_3, p_4) = p_2 \oplus p_3$.  A basic truth table shows that the function pattern is 0011, 1100,  0011, 1100 or $3C3C in hexadecimal (base 16 code).  This is function g in Table 4.  The result may be verified in Table 5.

**Design Problem B** – The input $|x,1\rangle$ is given as $|0,0,0,0,1\rangle$ and the function is given as $3333.  Determine the resulting output.

**Solution B** – Note that $3333 means the truth table pattern is:  0011, 0011, 0011, 0011.  It is soon found that $f(p_1, p_2, p_3, p_4) = p_2$ delivers this table.  This is function e in Table 4.  Therefore, the output state is $|0, 0, 1\ 0, 1\rangle$ as may be seen from a wiring diagram applied as above.  Table 5 verifies the result.

**ENCRYPTED MESSAGES**

The test system in Figure 1 begins with what may be described as a particular basis state of dimension n+1, that is $|0, 0, 0,\ldots,1\rangle$..  Then a Hadamard transform of order n+1 creates alternating positive and negative entries.  The state vector is then multiplied by a certain $U_f$ representing a certain quantum Boolean function, thus mixing up the elements of the state vector.  The physical system with the mixed up state vector is subsequently sent to another person, the receiver, who does not know the function and the basic qubits being sent.  Can the original qubits and the Boolean function be identified with any degree of efficiency?

Essentially the receiver must know enough to multiply the received vector by $\mathbf{R^{n+1}} = \mathbf{H} \otimes \mathbf{H} \otimes \ldots \otimes \mathbf{H}$ and then observe individual qubits. Once the state vector is observed, it is permanently modified and cannot be observed again.

In general, there will be a chance of any possible basis state since probability could be distributed over each element in the state vector. So a given observation does not necessarily define the original state vector. To circumvent uncertainty, it usually is agreed apriori that the function is a member of one of the following classes.

**Class 1, constant quantum function** -- The resulting observation of n+1 qubits could be |0>|0>|1> implying a state vector of the form $[0 \pm 1\ 0\ 0\ \ldots 0]^T$. This is a basis state with the second element finite and all others zero. One will always observe |0>|0>|1> with 100 % probability. In this case, the function is definitely a <u>constant</u>. A plus sign means a constant function that is 0 for all possible binary inputs in a truth table. A negative sign means a constant function that is 1 for all possible binary inputs in a truth table. A negative sign will not be observable however.

**Class 2, symmetric or anti-symmetric quantum function** – These functions follow from symmetric or antisymmetric truth tables of dimension $2^{n+1}$. Therefore $2^n$ codes are determined with 100 % probability. A constant function is a special case of this.

**Class 3, balanced function** – A function could be balanced, meaning equal numbers of ones and zeros in a truth table for the Boolean function. For example, the function with the truth table 01110001 is balanced, although it is neither symmetric nor anti-symmetric. A balanced function can be shown to have a state vector of the form $[x_0\ 0\ x_1\ x_2\ \ldots x_{n+1}]'$. That is, the coefficient for the basis state |000…01> will necessarily be zero. The remaining basis states in this vector may or may not have finite coefficients. So if a measurement yields qubits |0>|0>…|0>|1> the subject function is certainly is not a balanced function because there is zero probability of getting this.

Returning to the communications issue, it may be agreed that only state vector systems of class 1 or 3 are going to be used, either constant or balanced. This relates to the Deutsch and the Deutsch-Jozsa algorithm. With this a priori knowledge, a single measurement would identify the function type with certainty. If constant a receiver must obtain qubits |0>|0>…|0>|1>; if balanced a receiver will obtain something else, anything but this.

Alternatively, it may be agreed by sender and receiver that only symmetric or antisymmetric functions (class 2) are going to be used. With this a priori knowledge, a single measurement would identify with certainty, one of the $2^n$ quantum Boolean symmetric or antisymmetric functions (or one of the $2^n$ complements of these functions). So there are $2^n$ unique possibilities to a receiver any one of which may be observed with 100 % certainty with one measurement.

Classically there would have to be $2^n$ tests to determine the codes which make the function true, which is significant work for large n.

**CONCLUSIONS**
Symmetric and anti-symmetric functions are important because there are many such functions ($2^{n+1}$ for n qubits), and because the resulting qubit states can be observed experimentally with 100 % certainty.

Such functions may serve to exercise a quantum computer. If all is functional, the expected basis state is measured with 100 % certainty. If the system has lost coherence (or has a design problem) the expected result will not be certain.

Negative symmetric or antisymmetric functions are available to force multiplication of a state vector by –1, that is, to shift phase by ± 180 °.

If any two of three parameters, 1] vector input, 2] $U_f$, 3] vector output) are known, the other parameter can be determined as demonstrated above.

Symmetric or anti-symmetric functions could serve to encrypt binary information using a Deutsch-Jozsa style algorithm.

**APPENDIX 1**
**SYMMETRIC AND ANTI-SYMMETRIC FUNCTION CONSTRUCTION**
Symmetric and anti-symmetric functions may be built from the bottom up beginning with n = 1. For n = 1, one bit may be either 0 or 1, denoted as a list of two items 0, 1. The first item is concatenated with itself and with its mirror image in the row, in this case, the second item. This creates 00, 01. These are interpreted as positive functions or truth tables. The leftmost bit denotes the polarity of the function (0 = positive). To obtain negative functions, the last item in the row is concatenated with itself, and with its mirror image in the row, in this case, the first item. This creates two more functions 11, 10 and judging by the leftmost bit they are negative functions (1 = negative).

To obtain the functions <u>for n = 2</u>, we place the functions for n = 1 in numerical order: 00, 01, 10, 11. There are four items in this list. The general method is to concatenate each item with itself and then with its mirror image in the list: the 4 positive functions are 0000, 0011, 0101, 0110. The 4 negative functions are 1111, 1100, 1010, 1001. These occur in reverse numerical order. The resulting functions may now be expressed in numerical order: 0000, 0011, 0101, 0110, 1001, 1010, 1100, 1111.

For n = 3, we follow the general method. Concatenate each item in the list with itself and then with its mirror image in the list. The 8 positive functions are 00000000, 00001111, 00110011, 00111100, 01010101, 01011010, 01100110, and 01101001; the 8 negative functions occur in reverse numerical order but may be placed in numerical order: 10010110, 10011001, 10100101, 10101010, 11000011, 11001100, 11110000, and 11111111.

In general for n bits, there will be $2^n$ functions of the positive type, and another $2^n$ of the negative type, or a total of $2^{n+1}$. The number of functions grows by factors of two since a list of $2^n$ functions results in $2^{n+1}$ functions. Each is either symmetric or anti-symmetric by virtue of its construction.

**APPENDIX 2**
**Table 5. Partial Conversion Table (n = 4); Inputs on Top; Outputs on Left. Function Patterns a, b, c, …o Identified In Table 4.**

| \|x,1>   \|y,1> | **00001** | 00011 | 00101 | 00111 | 01001 | 01011 | 01101 | 01111 | **10001** | 10011 | 10101 | 10111 |
|---|---|---|---|---|---|---|---|---|---|---|---|---|
| 00001 00011 | a i | i a | e m | m e | c k | k c | g o | o g | b j | j b | f n | n f |
| **00101** 00111 | e m | m e | a i | i a | g o | o g | c k | k c | f n | n f | b j | j b |
| 01001 01011 | c k | k c | g o | o g | a i | i a | e m | m e | d l | l d | h p | p h |
| 01101 01111 | g o | o g | c k | k c | e m | m e | a i | i a | h p | p h | d l | l d |
| 10001 10011 | b j | j b | f n | n f | d l | l d | h p | p h | a i | i a | e m | m e |
| 10101 10111 | f n | n f | b j | j b | h p | p h | d l | l d | e m | m e | a i | i a |
| 11001 11011 | d l | l d | h p | p h | b j | j b | f n | n f | c k | k c | g o | o g |
| **11101** 11111 | h p | p h | d l | l d | f n | n f | b j | j b | g o | o g | c k | k c |